\documentclass[twocolumn]{aa}

  \usepackage{amscd}
  \usepackage{amsmath}
  \usepackage{amssymb}
  \usepackage[T1]{fontenc}
  \usepackage[varg]{txfonts}
  \usepackage{graphicx}
  \usepackage{natbib}
  \usepackage{verbatim}
  \usepackage{array}

\begin{document}

\title{The distributed burning regime in Type Ia supernova models}

\author{F. K. R{\"o}pke
        \and
        W. Hillebrandt}
          
   \offprints{F. K. R{\"o}pke}

   \institute{Max-Planck-Institut f\"ur Astrophysik,
              Karl-Schwarzschild-Str. 1, D-85741 Garching, Germany\\
              \email{fritz;wfh@mpa-garching.mpg.de}
             }

\abstract{The deflagration mode of flame propagation in Type Ia
  supernova (SN Ia) models requires a correct description of the
  interaction of the flame with turbulent motions. It is well-known that
  turbulent combustion proceeds in different regimes. For most parts of the
  deflagration in SNe Ia the flamelet regime
  applies. This has been modeled in previous multi-dimensional
  simulations. However, at late stages of the explosion, the
  flame will propagate in the so-called distributed regime. We
  investigate the effects of this regime on SN Ia models in a first
  and simplified approach and show that the trend of effects seems
  capable of curing some problems of current pure deflagration
  models.
}

\maketitle


\section{Introduction}

\cite{damkoehler1940a} was the first to distinguish between different
regimes of turbulent combustion. Since his pioneering work progress in
the theoretical understanding of the turbulent combustion process
\citep[e.g.][]{peters2000a}
as well as flame experiments in the laboratory
\citep[see][for a collection of data]{abdel-gayed1981a} have confirmed
his ideas. He described a regime of 
``large-scale turbulence'' where the flame is deformed by interaction
with turbulent motions and a ``small-scale turbulence'' regime where
turbulent eddies actually penetrate the internal flame structure. In modern
perception these regimes are identified with the \emph{flamelet} and the
\emph{distributed} regimes, respectively.

As pointed out by \citet{niemeyer1997b}, both regimes should be
reached subsequently in the deflagration model of SN Ia explosions
\citep[for a review of models
see][]{hillebrandt2000a}. In a statistical approach the distributed
burning regime was addressed in the context of SNe Ia by
\citet{lisewski2000a}. Direct numerical simulations of flames in
degenerate matter have recently been able to reach the distributed
burning regime \citep{bell2004b}.
Nonetheless, the distributed burning regime has never been implemented
in global deflagration models of thermonuclear supernovae
\citep[e.g.][]{reinecke2002b, gamezo2003a}. This was motivated by the
fact that due to the low fuel densities to which the regime applies,
no additional iron group nuclei are synthesized. Moreover, it was assumed that
the flame propagation here would be too slow to 
significantly contributing to the energy generation. This may be a
reasonable approach as long as the models aim at first order effects
of the explosion characteristics. However, constant development in
modeling and increasing computational resources facilitate systematic
tests of initial parameters \citep[see e.g.][]{roepke2004c,roepke2004e} and
first synthetic light curves were derived by
\citet{sorokina2003a}. Even synthetic spectra may be calculated
from multi-dimensional models soon.

Here the question arises whether the pure deflagration models (in
which the flame propagates subsonically) in their
current form are consistent with observational data. The
physics input of these models may be incomplete. 
Possible problems arise from their low explosion energies which may
only reproduce the low side of the average observed velocities of the
ejecta in SNe Ia. Furthermore, 
they seem to underproduce intermediate mass
elements and leave unburnt material at low velocities.
Based on their success
in one-dimensional parametrized models, delayed detonations
have been put forward as a favorable scenario \citep{gamezo2004a}. In
a detonation the flame is mediated by shock waves and travels with
sound speed. The problem with the delayed detonation scenario is that
no physical mechanism could be identified yet that would trigger the
transition from a deflagration to a detonation under conditions of SN
Ia explosions \citep{niemeyer1999a,roepke2004a,roepke2004b}. 

With reconsidering the distributed burning regime in the deflagration
model of thermonuclear supernovae we propose an alternative mechanism
that may help to overcome some of the difficulties of that model without
artificially evoking a detonation. The advantage of this extended
deflagration model is that it rests on a sound basis of known
physics. If no other effects terminate the deflagration phase, burning
inevitably enters the distributed regime.

Here, we study the effects of the distributed burning in very
simplified two-dimensional models. As will be 
discussed below the employed description of this regime is no more
than a coarse first-order estimate and can only point out the trends
how it
would change the model. Substantially more effort will be needed to
implement details of the distributed burning regime in global SN Ia
explosion simulations. 

\section{Modeling turbulent combustion in SNe Ia}

The propagation speed of a laminar deflagration flame is determined by
a balance between energy production in the reactions and diffusive energy
transport. Based on this simple idea, \cite{mikhelson1889a} derived an
expression for the laminar burning velocity. This expression was later
corrected by \citet{zeldovich1938a} taking into account the activation
energy necessary 
to induce the reaction. This results in a division of the flame
structure into a preheat zone and an reaction zone. In both
theories, the laminar flame speed $s_\mathrm{l}$ turns out to be
proportional to the square root of the diffusivity $D$.

The wrinkling of the thermonuclear flame in SN Ia explosions begins at
large scales. An inverse stratification of light
ashes below dense fuel in the gravitational field of the exploding
white dwarf (WD) star makes burning from its center outward
intrinsically unstable. Buoyancy (Rayleigh-Taylor) instabilities lead
to the formation of burning bubbles that rise into the cold fuel. This
effect is directly resolved in multidimensional SN Ia
simulations. Secondary shear (Kelvin-Helmholtz) instabilities at the
interfaces of the burning bubbles give rise to the generation of
turbulent motions. The formed turbulent eddies decay to smaller scales
in a turbulent cascade.

The eddies of the turbulent cascade interact with the flame further
wrinkling and stretching it on smaller scales. This leads to a rapid
increase in the flame surface and results in an enhanced burning
rate. This effect acts down to the so-called Gibson scale
\citep{peters2000a}. Below that scale turbulent velocities are so
small compared with the laminar burning speed of the flame that the
flame burns through turbulent eddies before they can significantly
alter its shape. If the Gibson length is large compared with the width
of the flame, turbulent burning proceeds in the
\emph{flamelet regime}. Here the flame is deformed by turbulence, but
its internal structure is not affected.

In the flamelet regime, the effect of surface enlargement on
numerically unresolved small scales can be compensated by attributing
an adequate effective turbulent burning velocity $s_\mathrm{t}$ to a (resolved)
smoothed flame front. This effective turbulent burning velocity has to
be determined in a way that the mass flux through the smoothed flame
equals that through the unsmoothed flame propagating with its laminar
speed. \citet{damkoehler1940a} found that in the flamelet regime the turbulent
flame velocity completely decouples from the laminar speed and is
proportional to the turbulent velocity fluctuations $v'$:
\begin{equation}\label{flamelet_eq}
s_\mathrm{t} \propto v'
\end{equation}

Due to the expansion in the explosion process, the fuel density drops
and this broadens the flame width \citep[cf.][]{timmes1992a}. At some stage
the turbulent motions become capable of penetrating the internal flame
structure. The flame
enters the \emph{distributed burning
  regime} \citep[see][for an analysis of this transition in SNe
Ia]{niemeyer1997d}. Damk{\"o}hler's idea to describe this regime was to assume
that the turbulent eddies entering the preheat zone alter the
diffusivity by mixing its thermal structure. Therefore, in analogy to
the expression for the laminar
burning velocity, the turbulent burning velocity in the distributed
regime should scale like
$$
s_\mathrm{t} / s_\mathrm{l} \sim
\left(D_\mathrm{t}/D\right)^{1/2},
$$
with $D_\mathrm{t}$ denoting a \emph{turbulent diffusivity}. The
laminar diffusivity $D$ is proportional to product of the flame thickness
$l_\mathrm{f}$ and the laminar flame speed
$s_\mathrm{l}$. Analogously, $D_\mathrm{t}$ is set proportional to the
turbulent velocity fluctuations $v'$ at a certain length scale
$l$. For the turbulent flame speed in the distributed
regime this yields a square-root dependency on $v'$,
\begin{equation}\label{distributed_eq}
s_\mathrm{t} \sim \left(s_\mathrm{l}\, v'\, l / l_\mathrm{f}\right)^{1/2},
\end{equation}
which is confirmed by experimental data. In this regime $s_\mathrm{t}$
scales with the laminar flame speed. Turbulent mixing of the
internal structure of the flame accelerates its propagation velocity.


To model both turbulent burning regimes in numerical simulations of
SNe Ia we apply the scheme proposed by
\citet{reinecke1999a,reinecke2002b} and \citet{roepke2004d}. The
hydrodynamics is solved for by the piecewise parabolic method on a
computational grid co-expanding with the WD star and the
flame is modeled using the level-set approach. In the latter, the
effective burning velocity of the flame has to be provided.
To this end $v'$ is derived from a sub-grid scale model
\citep{niemeyer1995b}. The maximum value of $s_\mathrm{l}$ and
expression (\ref{flamelet_eq}) then gives the flame propagation velocity in
the first stages of the explosion process.

\begin{figure}[t]
\centerline{
\includegraphics[width = 0.8 \linewidth]
  {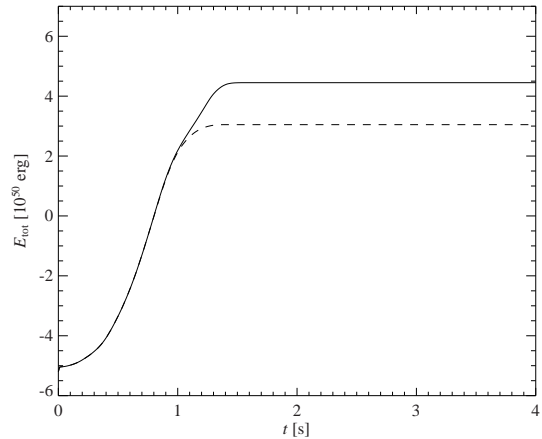}
}
\caption{Total energies released by the model ignoring
  distributed burning (dashed) and the model including the
  distributed burning phase (solid) . \label{etot_fig}}
\end{figure}

Previous models stopped burning in the flame once the 
fuel density fell below $10^7 \, \mathrm{g} \,
\mathrm{cm}^{-3}$. To take into account the
distributed regime we extend this burning and and derive
$s_\mathrm{t}$ form Eq.~(\ref{distributed_eq}) at densities below
$10^7 \, \mathrm{g} \, \mathrm{cm}^{-3}$ down to $5 \times 10^5 \,
  \mathrm{g} \, \mathrm{cm}^{-3}$.
The necessary values for the laminar flame speeds $s_\mathrm{l}$ and
the flame thickness $l_\mathrm{f}$ at low fuel densities are provided
by fits to the data given by \citet{bell2004a,bell2004b}. The
turbulent velocity fluctuations $v'$ on the
computational grid scale $l$ are derived from the sub-grid scale
model.

The feasibility of modeling the distributed burning regime via the level
set approach -- though in a more elaborate way of formulating the flame
speed -- has been shown and validated against experiments by
\citet{duchamp2002a}.

\begin{figure}[t]
\centerline{
\includegraphics[width = 0.84 \linewidth]
  {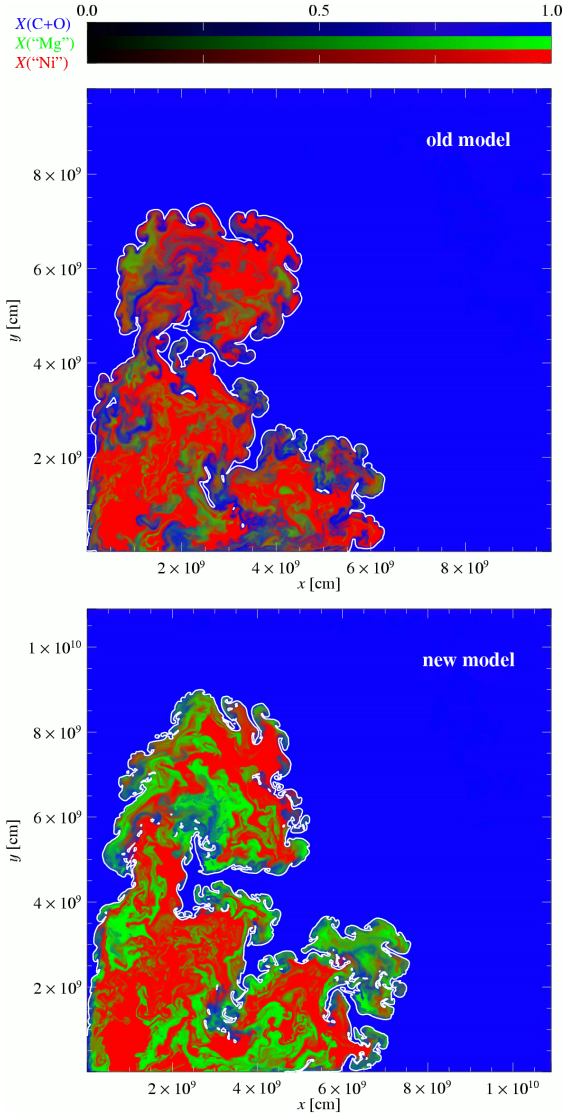}
}
\caption{Snapshots of the models with $[1024]^2$ cells at $t = 10 \, \mathrm{s}$. \label{comp_fig}}
\end{figure}

\section{Results of numerical models}

We compare two very simple two-dimensional SN Ia models calculated on
a $1024\times 1024$ cells grid: one ignoring
the distributed burning (referred to as ``old'' model) and one taking into
account this regime (``new'' model). In both
cases the flame was ignited in the \emph{c3} shape of
\citet{reinecke2002b} -- a central ignition superposed by sinusoidal
perturbations. The initial resolution of both models was $1.975 \,
\mathrm{km}$. Due to the different energy releases the final
resolutions were $96 \, \mathrm{km}$ for the old and
$100\,\mathrm{km}$ for the new model.

The total energy release of both models is plotted in
Fig.~\ref{etot_fig}, which also includes results of a resolution
  study of the new model showing convergence for more than $[512]^2$
  cells. Contrary to more sophisticated three-dimensional
models the simulations discussed here are based on very
simple two-dimensional setups and they cannot be expected to reproduce
the explosion strength of observed SNe Ia. Nevertheless, trends can be
inferred even from these simple models. Clearly, the distributed
regime contributes
substantially to the energy generation. About $1.1 \,\mathrm{s}$ after
ignition the energy generation in the old model
ceases because the fuel density reaches low values. In the new model,
however, the burning proceeds for another $0.3 \,
\mathrm{s}$, leading to a $\sim$17\% increase in the nuclear energy
release.
In this time range, the distributed burning
produces large amounts of intermediate mass elements.

\begin{figure}[t]
\centerline{
\includegraphics[width = 0.8 \linewidth]
  {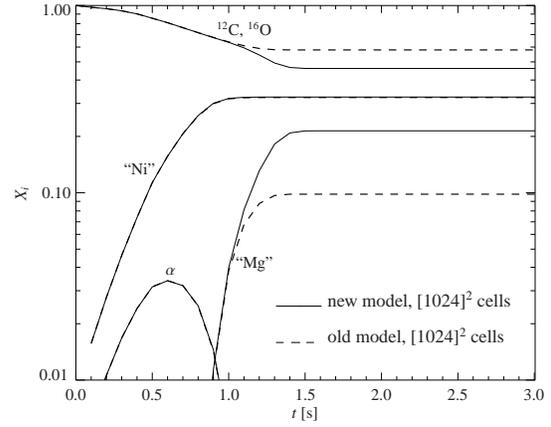}
}
\caption{Temporal evolution of the chemical composition in the
  models.\label{evalcomp_fig}}
\end{figure}

\begin{figure}[t]
\centerline{
\includegraphics[width = 0.8 \linewidth]
  {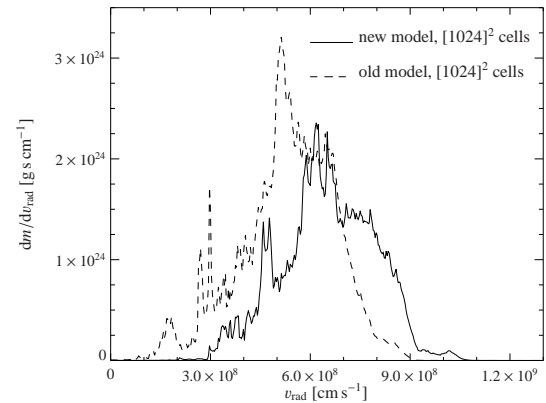}
}
\caption{Distribution of unburnt material in velocity space after
  $10\,\mathrm{s}$.\label{vprof_fig}}
\end{figure}

This is evident by comparing the snapshots of both simulations at $t =
10 \, \mathrm{s}$ shown in Fig.~\ref{comp_fig}. They display the position
of the zero level set as white curve \citep[indicating the position
of the flame front while
burning is active and still providing an approximate separation
between fuel and ashes later in the simulations; see][]{roepke2004d}.
The distribution of the species is color-coded. Our simplified
description of the nuclear burning \citep[for details
see][]{reinecke2002b} pools the intermediate mass nuclei in the
representative element ``Mg''  and represents the iron group nuclei by
``Ni''. The evolution of the chemical composition in both models is
compared in Fig.~\ref{evalcomp_fig}.

Implementing the distributed burning regime into our simulations we
note various important aspects of the results. 
Due to the larger energy release the ejecta reach higher velocities
and expand faster. This is obvious from Fig.~\ref{comp_fig}.
The transition from
burning in the flamelet regime to the distributed regime takes place
at low fuel densities (and is implemented in our models this
way). Therefore the new and old models differ only in the production
of intermediate mass nuclei. The amount of synthesized iron group
elements is unaffected (cf.\ Fig.~\ref{evalcomp_fig}). Moreover,
Fig.~\ref{comp_fig} shows that the
distribution of ``Ni''-rich regions in the new model resembles that in
the old one. However, in the new model these regions are embedded in a
layer of intermediate mass elements, which also shows a turbulent
structure.

Taking into account the distributed regime, the flame naturally
consumes more fuel. Therefore, the unburnt material in the
central parts is strongly depleted. This is corroborated by
Fig.~\ref{vprof_fig} showing an angular average over the distribution
of unburnt material in velocity space. No unburnt material is left at
low velocities.

\section{Conclusions}

By reconsidering the distributed burning regime in global SN Ia
models we have proposed a way to alleviate some of the problems that
might arise when comparing the deflagration model with observations.
By extending the burning of material to intermediate mass elements it
acts against a possible underproduction of those in previous
deflagration models. The additional energy released here helps to
accelerate the ejecta to higher velocities. Near the center the
burning consumes now most of the fuel leaving no low-velocity carbon
and oxygen behind.

The newly implemented burning regime does, however, not affect the
production of iron group elements since it applies to low fuel
densities. It was claimed that to reproduce the observed spectra a
layered composition of the ejecta would be needed
\citep{gamezo2004a}. This is, however, not yet confirmed since
synthetic spectra derived with multi-dimensional radiation transport
schemes are still lacking. Recent spectrapolarimetry data
\citep{wang2004a}
seem to indicate that a turbulent structure of the intermediate mass
elements is required. In any case, the deflagration phase does not
remove the mixed distribution of species in the ejecta that was
present already in the old models.

The results presented here have to be considered preliminary for
several reasons. Our simulations are performed
only in two spatial dimensions. The turbulent structure develops
differently in three-dimensional models resulting in an overall more
vigorous explosion \citep[e.g.][]{reinecke2002b,roepke2004d}. An even
more important issue is that the modeling of the distributed burning
regime applied here is very crude and no more than a first-order
approach. The value of transition density is an \emph{ad hoc} choice.
Moreover, Eq.~(\ref{distributed_eq}) is in a strict sense a limiting
case for $l_\mathrm{f} \gg l$, which is certainly not yet met at the
presumed transition density. Therefore our description by simply
switching from the flamelet
regime to the distributed regime will overestimate the
effects. 

However, the presented study points to the trends that can be
expected. The distributed burning regime has to be implemented into global
SN Ia simulations to complete the deflagration model. Important
consequences for the energetics and the derived synthetic spectra may
result from this extension, although the effects may not be as large
as in the crude first models presented here. Whether they suffice to
remove the current problems of the deflagration
scenario will be investigated in forthcoming studies. For instance, a
more
general description of the turbulent flame propagation velocity
\citep[featuring a smooth transition between both burning
regimes; see][]{peters2000a} will be tested and also applied to
three-dimensional models.

\begin{acknowledgements}
We thank the Institute for Nuclear Theory at the University of
Washington for its hospitality.
\end{acknowledgements}




\begin{thebibliography}{26}
\expandafter\ifx\csname natexlab\endcsname\relax\def\natexlab#1{#1}\fi

\bibitem[{{Abdel-Gayed} \& {Bradley}(1981)}]{abdel-gayed1981a}
{Abdel-Gayed}, R.~G. \& {Bradley}, D. 1981, Phil. Trans. R. Soc. Lond. A, 301,
  1

\bibitem[{{Bell} {et~al.}(2004{\natexlab{a}}){Bell}, {Day}, {Rendleman},
  {Woosley}, \& {Zingale}}]{bell2004a}
{Bell}, J.~B., {Day}, M.~S., {Rendleman}, C.~A., {Woosley}, S.~E., \&
  {Zingale}, M. 2004{\natexlab{a}}, \apj, 606, 1029

\bibitem[{{Bell} {et~al.}(2004{\natexlab{b}}){Bell}, {Day}, {Rendleman},
  {Woosley}, \& {Zingale}}]{bell2004b}
{Bell}, J.~B., {Day}, M.~S., {Rendleman}, C.~A., {Woosley}, S.~E., \&
  {Zingale}, M. 2004{\natexlab{b}}, \apj, 608, 883

\bibitem[{{Damk{\"o}hler}(1940)}]{damkoehler1940a}
{Damk{\"o}hler}, G. 1940, Z. f. Elektroch., 46, 601

\bibitem[{{Duchamp de Langeneste} \& {Pitsch}(2002)}]{duchamp2002a}
{Duchamp de Langeneste}, L. \& {Pitsch}, H. 2002, in Annual Research Briefs
  (Center for Turbulence Research), 91--101

\bibitem[{{Gamezo} {et~al.}(2004){Gamezo}, {Khokhlov}, \& {Oran}}]{gamezo2004a}
{Gamezo}, V.~N., {Khokhlov}, A.~M., \& {Oran}, E.~S. 2004, \prl, 92, 211102

\bibitem[{{Gamezo} {et~al.}(2003){Gamezo}, {Khokhlov}, {Oran}, {Chtchelkanova},
  \& {Rosenberg}}]{gamezo2003a}
{Gamezo}, V.~N., {Khokhlov}, A.~M., {Oran}, E.~S., {Chtchelkanova}, A.~Y., \&
  {Rosenberg}, R.~O. 2003, Science, 299, 77

\bibitem[{{Hillebrandt} \& {Niemeyer}(2000)}]{hillebrandt2000a}
{Hillebrandt}, W. \& {Niemeyer}, J.~C. 2000, \araa, 38, 191

\bibitem[{{Lisewski} {et~al.}(2000){Lisewski}, {Hillebrandt}, {Woosley},
  {Niemeyer}, \& {Kerstein}}]{lisewski2000a}
{Lisewski}, A.~M., {Hillebrandt}, W., {Woosley}, S.~E., {Niemeyer}, J.~C., \&
  {Kerstein}, A.~R. 2000, \apj, 537, 405

\bibitem[{{Mikhel'son}(1889)}]{mikhelson1889a}
{Mikhel'son}, V.~A. 1889, PhD thesis, Univ. Moscow, see {\em Collected works},
  Vol.~1, Novyi Agronom Press, Moscow (1930)

\bibitem[{{Niemeyer}(1999)}]{niemeyer1999a}
{Niemeyer}, J.~C. 1999, \apj, 523, L57

\bibitem[{{Niemeyer} \& {Hillebrandt}(1995)}]{niemeyer1995b}
{Niemeyer}, J.~C. \& {Hillebrandt}, W. 1995, \apj, 452, 769

\bibitem[{{Niemeyer} \& {Kerstein}(1997)}]{niemeyer1997d}
{Niemeyer}, J.~C. \& {Kerstein}, A.~R. 1997, New Astronomy, 2, 239

\bibitem[{{Niemeyer} \& {Woosley}(1997)}]{niemeyer1997b}
{Niemeyer}, J.~C. \& {Woosley}, S.~E. 1997, \apj, 475, 740

\bibitem[{{Peters}(2000)}]{peters2000a}
{Peters}, N. 2000, Turbulent Combustion (Cambridge: Cambridge University Press)

\bibitem[{{Reinecke} {et~al.}(1999){Reinecke}, {Hillebrandt}, {Niemeyer},
  {Klein}, \& {Gr{\" o}bl}}]{reinecke1999a}
{Reinecke}, M., {Hillebrandt}, W., {Niemeyer}, J.~C., {Klein}, R., \& {Gr{\"
  o}bl}, A. 1999, \aap, 347, 724

\bibitem[{{R{\" o}pke} \& {Hillebrandt}(2004{\natexlab{a}})}]{roepke2004e}
{R{\" o}pke}, F.~K. \& {Hillebrandt}, W. 2004{\natexlab{a}}, A\&A in
print, astro-ph/0409286

\bibitem[{{R{\" o}pke} \& {Hillebrandt}(2004{\natexlab{b}})}]{roepke2004c}
{R{\" o}pke}, F.~K. \& {Hillebrandt}, W. 2004{\natexlab{b}}, \aap, 420, L1

\bibitem[{{R{\" o}pke} {et~al.}(2004{\natexlab{a}}){R{\" o}pke}, {Hillebrandt},
  \& {Niemeyer}}]{roepke2004a}
{R{\" o}pke}, F.~K., {Hillebrandt}, W., \& {Niemeyer}, J.~C.
  2004{\natexlab{a}}, \aap, 420, 411

\bibitem[{{R{\" o}pke} {et~al.}(2004{\natexlab{b}}){R{\" o}pke}, {Hillebrandt},
  \& {Niemeyer}}]{roepke2004b}
{R{\" o}pke}, F.~K., {Hillebrandt}, W., \& {Niemeyer}, J.~C.
  2004{\natexlab{b}}, \aap, 421, 783

\bibitem[{{R{\"o}pke}(2004)}]{roepke2004d}
{R{\"o}pke}, F.~K. 2004, submitted to A\&A, astro-ph/0408296

\bibitem[{{Reinecke} {et~al.}(2002){Reinecke}, {Hillebrandt}, \&
  {Niemeyer}}]{reinecke2002b}
{Reinecke}, M., {Hillebrandt}, W., \& {Niemeyer}, J.~C. 2002, \aap, 386, 936

\bibitem[{{Sorokina} \& {Blinnikov}(2003)}]{sorokina2003a}
{Sorokina}, E. \& {Blinnikov}, S. 2003, in From Twilight to Highlight: The
  Physics of Supernovae, ed. W.~{Hillebrandt} \& B.~{Leibundgut}, ESO
  Astrophysics Symposia (Berlin Heidelberg: Springer-Verlag), 268--275

\bibitem[{{Timmes} \& {Woosley}(1992)}]{timmes1992a}
{Timmes}, F.~X. \& {Woosley}, S.~E. 1992, \apj, 396

\bibitem[{Wang {et~al.}(2004)Wang, Baade, H{\"o}flich, Wheeler, Kawabata,
  Khokhlov, Nomoto, \& Patat}]{wang2004a}
Wang, L., Baade, D., H{\"o}flich, P., {et~al.} 2004, submitted to ApJ,
  astro-ph/0409593

\bibitem[{{Zel'dovich} \& Frank-Kamenetsky(1938)}]{zeldovich1938a}
{Zel'dovich}, Y.~B. \& Frank-Kamenetsky, D.~A. 1938, Acta Physicochim. URSS, 9,
  341

\end{thebibliography}

\end{document}